\documentclass[11pt]{article}
\usepackage{moriond,epsfig}

\def\be{\begin{equation}}
\def\ee{\end{equation}}
\def\bea{\begin{eqnarray}}
\def\eea{\end{eqnarray}}

\def\nn{\nonumber \\}
\usepackage{amssymb}
\begin{document}
\vspace*{4cm}
\title{YUKAWA UNIFICATION IN SUSY SO(10) FOR $\mathbf{\mu<0}$ \\ CONSISTENT WITH MUON $\mathbf{g-2}$ AND $\mathbf{b\to s\gamma}$}

\author{ MARCIN BADZIAK${}^{1,2}$ }

\address{${}^1$DAMTP, University of Cambridge, Wilberforce Road, Cambridge CB3 0WA, UK \\
${}^2$Cavendish Laboratory, University of Cambridge, J.J. Thomson Avenue, Cambridge CB3 0HE, UK
}

\maketitle\abstracts{
It is shown that top-bottom-tau Yukawa unification for $\mu<0$ can be consistent with $(g-2)_{\mu}$ and $b\to s\gamma$.  This happens for non-universal gaugino masses which are assumed to be generated by the F-term vev in a 54-dimensional representation of SO(10). The requirement of $(g-2)_{\mu}$ and $b\to s\gamma$ being within $2\sigma$ from the experimental central values, together with the correct relic abundance of neutralinos, leads to rather definite predictions for sparticle spectrum. In particular, the gluino mass is predicted to be between $500$ and $700$ GeV or between $900$ GeV and $1.6$ TeV. }

\section{Introduction}

Apparent gauge coupling unification in the Minimal Supersymmetric Standard Model (MSSM) is regarded as the one of the best motivations for supersymmetry (SUSY). Among the
candidates for the unified gauge group SO(10) seems to be the most attractive. All Standard Model (SM) matter fields, as well as right-handed neutrino, of each generation fit into one
irreducible 16-dimensional representation of SO(10). On the other hand, two MSSM Higgs doublets, $H_u$ and $H_d$, sit in the 10-dimensional representation. Moreover, in the
simplest version of SUSY SO(10) Yukawa couplings of top, bottom and tau unify at the same scale as the gauge couplings do.

One of the generic predictions of Yukawa unification is a large value of $\tan\beta\sim m_t/m_b$. For such values of $\tan\beta$ there are sizable threshold corrections to the bottom mass which are of major importance from the point of view of bottom-tau Yukawa unification. The main finite corrections, originating from gluino-sbottom and chargino-stop loops, are given by \cite{Hall,Carena,Pierce}:
\begin{equation}
\label{mbsusycorr}
\left(\frac{\delta m_b}{m_b}\right)^{\rm finite}
\approx\frac{g_3^2}{6\pi^2}\mu m_{\tilde g}\tan\beta \, 
I(m_{\tilde b_1}^2,m_{\tilde b_2}^2,m_{\tilde g}^2)
+\frac{h_t^2}{16\pi^2}\mu A_t\tan\beta \, 
I(m_{\tilde t_1}^2,m_{\tilde t_2}^2,\mu^2) \,,
\end{equation} 
where the loop integral $I(x,y,z)$ is well approximated by $a/\max(x,y,z)$ with $a$ between $0.5$ and $1$. Bottom-tau Yukawa unification requires the above correction to be negative with the magnitude about 10\% to 20\% \cite{Wells_yuk}. The gluino-sbottom correction dominates over the most of the parameter space so Yukawa unification generically prefers $\mu<0$. 

The sign of $\mu$ has also crucial impact on the muon anomalous 
magnetic moment, $(g-2)_{\mu}$, which experimental value is more than $3\sigma$ below the Standard Model prediction. The sign of the dominant SUSY contribution to $(g-2)_{\mu}$ is the same as the sign of the product $\mu M_2$. So, in phenomenologically acceptable models with  negative $\mu$ gaugino masses with $M_2<0$ are required.

In GUT theories gaugino masses are usually assumed to be equal at the GUT scale. Under this assumption $(g-2)_{\mu}$ calls for $\mu>0$ which makes top-bottom-tau Yukawa unification very problematic. However, gaugino masses are universal only if the SUSY breaking $F$-term which gets a VEV is a singlet of the GUT gauge symmetry group. In general, the gaugino masses in supergravity can arise from the following dimension five operator:
\begin{equation}
{\cal L}\supset -\frac{F^{ab}}{2 M_{\rm Planck}}
\lambda^a \lambda^b + {\rm c.c.}\,,
\end{equation}
where $\lambda^a$ are the gaugino fields. The vacuum 
expectation value of the relevant $F$-term, $\langle F^{ab}\rangle$, 
must transform as the singlet of the SM gauge group but  
it can be a non-singlet of the full GUT group. Since the gauginos 
belong to the adjoint representation, non-zero gaugino masses may 
arise from VEVs of the $F$-terms transforming as any of the 
representations present in the symmetric part of the direct 
product of the two adjoints, which for SO(10) is $({\bf 45} \times {\bf 45} )_S 
= {\bf 1} + {\bf 54} + {\bf 210} + {\bf 770}$. 

In this work we consider $\mu<0$ and assume that gaugino masses are generated by the $F$-term VEV transforming as ${\bf 54}$ representation. In such a case gaugino masses are given by \cite{Martin}:
\begin{equation}
\label{gauginoratio}
 (M_1,M_2,M_3) = \left(-\frac{1}{2},-\frac{3}{2},1\right)m_{1/2} \,.
\end{equation}

Tob-bottom-tau Yukawa unification requires also non-universal scalar masses to be compatible with radiative electroweak symmetry breaking \cite{Olechowski}. We assume the following pattern of scalar masses:
\begin{eqnarray}
m_{H_d}^2&=&m_{10}^2+2D\,, \nn[1pt]
m_{H_u}^2&=&m_{10}^2-2D\,, \nn[1pt]
m_{Q,U,E}^2&=&m_{16}^2+D\,, \nn[1pt]
m_{D,L}^2&=&m_{16}^2-3D\,. 
\end{eqnarray} 
$D$-term contribution which splits masses of scalars belonging to the same representation of SO(10) is a generic feature of models in which the GUT gauge group has larger rank than the SM gauge group \cite{Dterm}. The remaining free parameters in our model are the universal trilinear term $A_0$ and $\tan\beta$. 

In these proceedings we show that in the above setup top-bottom-tau Yukawa unification can be realized. Moreover, we present Yukawa-unified solutions which predict the values of BR$(b\to s \gamma)$ and $(g-2)_{\mu}$ consistent with the experimental data at $2\sigma$ level. Combination of these constraints, together with the WMAP bound for the relic density of neutralinos, imply that the model predicts light SUSY spectrum with no sparticles with masses above $2$ TeV.

\section{Necessary conditions for Yukawa unification}

Assuming that the finite threshold correction to the bottom mass is fully determined by the gluino-sbottom contribution (which is a good approximation over the most of the parameter space) the condition of top-bottom-tau leads to the upper bound for the parameter $|\mu|$ \cite{bop}:
\begin{equation}
 \label{mubound}
|\mu|\lesssim 0.4 m_{\tilde g}\approx m_{1/2}\,.
\end{equation}
In principle, the above bound for $|\mu|$ could be relaxed  if $m_{1/2}\ll m_{16}$. However, Yukawa-unified solutions respecting the hierarchy $m_{1/2}\ll m_{16}$ which are compatible with all experimental constraints cannot exist for large values of $|\mu/m_{1/2}|$ for the reason that we explain later on.

At large $\tan\beta$, the condition of proper REWSB implies $\mu^2\approx-\left(m_{H_u}^2+M_Z^2/2\right)$. Using this relation and the renormalization group equations one can estimate electroweak scale value of $\mu^2$ in terms of the input parameters at the GUT scale: 
\begin{equation}
\label{mucoeff_focus}
 \mu^2\approx -m_{H_u}^2\approx m_{1/2}^2\left[1.2 +0.65 x^2\left(0.97-\frac{m_{10}^2}{m_{16}^2}+2.9 \frac{D}{m_{16}^2}+0.15\frac{A_0^2}{m_{16}^2}-\frac{0.3}{x}\frac{A_0}{m_{16}}\right)\right] \,,
\end{equation}
where $x\equiv m_{16}/m_{1/2}$. In order to satisfy the bound (\ref{mubound}) the contribution from gaugino masses to $\mu^2$ has to be (partially) cancelled by other terms in (\ref{mucoeff_focus}). Yukawa unification requires also $D>0$ \cite{Murayama}. So, the cancellations in (\ref{mucoeff_focus}) may occur only for $m_{10}>m_{16}$. Since $\mu^2$ cannot be negative, Yukawa unification consistent with REWSB requires correlation between $m_{10}$, $D$ and $A_0$.
This correlation is especially strong when $m_{1/2}\ll m_{16}$ because in such a case the value of $\mu^2$ is very sensitive to the value of the expression in the round bracket in eq.\ (\ref{mucoeff_focus}).

\section{Interplay between BR$\mathbf{(b\to s \gamma)}$ and $\mathbf{(g-2)_{\mu}}$}

The main MSSM contribution to $(g-2)_{\mu}$ originates from the one-loop diagrams involving charginos accompanied by the muon sneutrino. Therefore, charginos and muon sneutrino have to be relatively light in order to explain $(g-2)_{\mu}$ anomaly. However, at large $\tan\beta$ light charginos give also significant contribution to BR$(b\to s \gamma)$ through the loops in which they are accompanied by up-type squarks. In order to avoid too large BR$(b\to s \gamma)$, the chargino contribution has to be necessarily negative, relative to the SM contribution.  

At large $\tan\beta$, there are two types of chargino contributions that may affect significantly the prediction for BR$(b\to s \gamma)$. The first one is proportional to stop-mixing angle and its sign is given by ${\rm sgn} \left(\mu A_t\right)$. The RG running gives negative $A_t$ with the absolute value of order $m_{1/2}$ unless $A_0$ is positive and few times larger than $m_{1/2}$ at the GUT scale.
This implies that for $\mu<0$ stop-mixing part of chargino contribution is typically positive. Fortunately, the sign of the second type of chargino contribution is given by $\left(-\mu M_2\right)$ which is always negative in our model. We call it gaugino contribution. This contribution is suppressed by squark GIM mechanism and vanishes for degenerate squark masses. In order to make chargino contribution to BR$(b\to s \gamma)$ negative, the gaugino contribution has to dominate over stop-mixing one. This is more likely when $m_{1/2}\ll m_{16}$ because in such a case intergenerational squark splitting may be large due to domination of RGEs by the terms proportional to Yukawa couplings. Moreover, $m_{1/2}\ll m_{16}$ typically suppresses stop-mixing angle. If there is no significant hierarchy between $m_{1/2}$ and $m_{16}$ chargino contribution is typically positive excluding the possibility of sizable SUSY contribution to $(g-2)_{\mu}$ unless $A_0$ is relatively large and positive.

\section{Yukawa-unified solutions consistent with $\mathbf{(g-2)_{\mu}}$ and $\mathbf{b\to s \gamma}$}

We performed numerical analysis using SOFTSUSY  \cite{softsusy} which solves 2-loop renormalization group equations implementing proper
REWSB and calculate sparticle spectrum. We also used MicrOmegas \cite{Micromega} for calculating the relic density of
dark matter, as well as, BR$(b\to s \gamma)$, $(g-2)_{\mu}$ and BR$(B_s\to\mu^+\mu^-)$.

Besides BR$(b\to s \gamma)$ and $(g-2)_{\mu}$, another important constraint for the model comes from the WMAP bound on the relic density of neutralinos. The requirement that these three constraints are satisfied leads to rather definite predictions for sparticle spectrum. In the following we discuss two types of Yukawa-unified solutions consistent with BR$(b\to s \gamma)$ and $(g-2)_{\mu}$ at $2\sigma$ level. 

In the first class of solutions there is no large hierarchy between $m_{1/2}$ and $m_{16}$ and $b\to s\gamma$ constraint is satisfied due to large positive $A_0$ at the GUT scale which allows for negative chargino contribution to BR$(b\to s \gamma)$.
In this class of solutions $m_{1/2}$ is found to be between about $400$ and $650$ GeV, while $m_{16}$ between $700$ and $1200$ GeV. In consequence, gluino mass is predicted to be between $900$ GeV and $1.6$ TeV. Squarks of the first and second generation are typically a bit heavier than the gluino. On the other hand, squarks of the third generation are typically a bit lighter than the gluino except $\tilde{b}_R$ which is much lighter due to negative $D$-term contribution at the GUT scale and strong renormalization by large $A_0$. In some cases $m_{\tilde{b}_R}$ is even below $200$ GeV which make annihilations of neutralinos through t-channel sbottom exchange efficient enough to satisfy the bound on $\Omega_{DM}h^2$. For larger values of $m_{\tilde{b}_R}$ this channel is less efficient and WMAP bound is satisfied due to neutralino co-annihilations with stau. 

We found also Yukawa-unified solutions which respect the hierarchy $m_{1/2}\ll m_{16}$. In this case correct relic abundance of neutralinos requires resonant annihilations through $Z$ boson or the lightest CP-even Higgs. This has a great impact on the allowed parameter space. First of all, these kind of annihilation channels are allowed only if the LSP has non-negligible higgsino component. This condition can be translated to the upper bound for $|\mu/M_1|$ which results also in the upper bound for $|\mu/m_{1/2}|$. We found numerically that in order to satisfy dark matter constraint $|\mu/m_{1/2}|$ has to be smaller than about $1.8$ and $1.5$ for $m_{\tilde{\chi}_1^0}\approx45$ GeV and $m_{\tilde{\chi}_1^0}\approx55$, respectively. For the LSP masses further away from the $Z$ or $h^0$ resonances $|\mu/m_{1/2}|$ has to be even smaller. This class of solutions predict gluino masses in the range between about $500$ and $700$ GeV. The squarks of the first and second generation have masses between $1.1$ and $1.5$ TeV, while the masses of third generation squarks are found to be between $800$ and $1000$ GeV except $\tilde{b}_R$ which has a mass between $400$ and $800$ GeV.

In summary: We have shown that top-bottom-tau Yukawa unification in SO(10) can be realized for $\mu<0$ in a way which is consistent with the experimental constraints from 
 $(g-2)_{\mu}$ and $b\to s \gamma$ if gaugino mases are generated by $F$-term VEV transfoming as 54-dimensional representation of SO(10). Moreover, for $\mu<0$ $D$-term splitting of scalar masses is compatible with Yukawa unification, in contrast to the case with positive $\mu$. This is the first SO(10) model which predicts light SUSY spectrum with all sparticle masses below $2$ TeV without violating any experimental constraints. The prediction of light gluinos make this model testable at the LHC in the very near future. More detailed analysis can be found in ref.\ \cite{bop}.

\section*{Acknowledgments}
M.B. would like to thank M.~Olechowski and S.~Pokorski for fruitful collaboration. This work has been partially supported by MNiSzW scientific research grant N N202 103838 (2010 - 2012) and by the European Research and Training Network (RTN) grant ‘Unification in the LHC era’ (PITN-GA-2009-237920).

\end{document}